\documentclass[journal,10pt]{IEEEtran}
\IEEEoverridecommandlockouts
\usepackage[T1]{fontenc}
\usepackage{algorithm}
\usepackage{algpseudocode}
\usepackage{amsmath}
\usepackage{amssymb}
\usepackage{amsfonts}
\usepackage{graphicx}
\usepackage{epsfig}
\usepackage{psfrag}
\usepackage{cite}
\usepackage{latexsym}
\usepackage{lettrine}
\usepackage{url}
\usepackage{color}
\usepackage{multirow}
\usepackage{subfigure}
\usepackage[justification=centering]{caption}
\usepackage{verbatim}

\DeclareMathOperator*{\argmax}{arg\,max}

\PassOptionsToPackage{bookmarks={false}}{hyperref}
\begin{document}
\captionsetup[figure]{name={Fig.},labelsep=period,singlelinecheck=off} 
\title{The Degrees-of-Freedom in Monostatic ISAC Channels: NLoS Exploitation vs. Reduction}

\author{Shihang Lu,~\IEEEmembership{Graduate Student Member,~IEEE}, Fan Liu,~\IEEEmembership{Member,~IEEE}, Lajos Hanzo,~\IEEEmembership{Life Fellow,~IEEE}
}

\maketitle
\begin{abstract}
The degrees of freedom (DoFs) attained in monostatic integrated
sensing and communications (ISAC) are analyzed. Specifically,
monostatic sensing aims for extracting target-orientation information
from the line of sight (LoS) channel between the transmitter and the
target, since the Non-LoS (NLoS) paths only contain clutter or
interference. By contrast, in wireless communications, typically, both
the LoS and NLoS paths are exploited for achieving diversity or
multiplexing gains. Hence, we shed light on the NLoS exploitation
vs. reduction tradeoffs in a monostatic ISAC scenario. In particular,
we optimize the transmit power of each signal path to maximize the
communication rate, while guaranteeing the sensing performance for the
target. The non-convex problem formulated is firstly solved in closed
form for a single-NLoS-link scenario, then we harness the popular
successive convex approximation (SCA) method for a general
multiple-NLoS-link scenario. Our simulation results characterize the
fundamental performance tradeoffs between sensing and communication,
demonstrating that the available DoFs in the ISAC channel should be
efficiently exploited in a way that is distinctly different from that
of communication-only scenarios.
\end{abstract}


\newtheorem{definition}{\underline{Definition}}[section]
\newtheorem{fact}{Fact}
\newtheorem{assumption}{Assumption}
\newtheorem{theorem}{\underline{Theorem}}[section]
\newtheorem{lemma}{\underline{Lemma}}[section]
\newtheorem{corollary}{\underline{Corollary}}[section]
\newtheorem{proposition}{\underline{Proposition}}[section]
\newtheorem{example}{\underline{Example}}[section]
\newtheorem{remark}{\underline{Remark}}[section]

\newcommand{\mv}[1]{\mbox{\boldmath{$ #1 $}}}
\setlength\abovedisplayskip{5pt}
\setlength\belowdisplayskip{5pt}

\section{Introduction}
In the forthcoming 5G-Advanced and 6G wireless networks, radio sensing
at the network level has been considered as a beneficial new feature
in support of emerging applications such as vehicle-to-everything
(V2X) communications, as well as smart city and unmanned aerial
vehicle (UAV) networks~\cite{cui2021integrating}. In the meantime,
given the imminent spectrum crunch of wireless communications, the
radar bands can be harnessed as alternative spectral resources. Owing
to the commonalities between sensing and communication (S$\&$C) in
terms of hardware architecture and signal processing
algorithms~\cite{liu2021integrated}, Integrated Sensing and
Communications (ISAC) constitutes a promising solution for embedding
attractive radar sensing functionalities into the existing cellular
infrastructure in a prompt and low-cost manner. Hence, it has received
tremendous research attention in the recent
years~\cite{liu2020Sensing-assistedComms,su2022secure,li2021rethinking,yang2020dual,zhang2022integrated,wang2021joint}.

Conventional radar systems are conceived solely for optimizing the
sensing performance without addressing the communication
functionality~\cite{fishler2006spatial,xu2008target,cui2013mimo}. With
the emerging integration of S$\&$C, both sensing-centric as well as
communication-centric and joint designs are proposed for ISAC signal
processing, which lead to flexible performance tradeoff between
S$\&$C~\cite{liu2021integrated}. In this spirit, novel resource
allocation and waveform design approaches were developed~\cite{liuxiang2020joint} for approaching the ISAC performance bounds,
leading to attractive integration and coordination
gains~\cite{liu2020Sensing-assistedComms}. More recently, numerous
design tradeoffs have been revealed~\cite{liu2021integrated}, ranging
from information theoretical tradeoffs to physical layer
tradeoffs~\cite{su2022secure}, and to cross-layer
designs~\cite{li2021rethinking}.

While the aforementioned contributions relied upon sophisticated
techniques, they generally assume that the S$\&$C signals propagate
over channels exhibiting similar statistical characteristics, even
though in practical scenarios, things become more complex. Briefly,
the presence of multiple paths can be exploited by the communication
functionality. But not all the paths are useful for radar
sensing~\cite{liu2021integrated}. Specifically, radar detection is
typically more dependent on the direct line of sight (LoS) link
between the radar transceiver and the target, since Non-LoS (NLoS)
links typically contain unwanted clutter~\cite{cui2013mimo}. By
contrast, wireless systems attain higher spatial degrees of freedom
(DoFs) by exploiting both LoS and NLoS links. By noting this
fundamental difference in S$\&$C propagation channels, it is of
pivotal importance to design efficient resource allocation for
exploiting the spatial DoFs inherent within the ISAC channels, which
motivates our research.

Explicitly, we consider an ISAC base station (ISAC-BS) that supports a
single communication user, which is also a target to be sensed. A
typical example for such a scenario is the sensing-assisted
vehicle-to-infrastructure (V2I) communication scenario, where a
roadside unit (RSU) wishes to communicate with a vehicle, while
simultaneously tracking its
movement~\cite{liu2020Sensing-assistedComms}. To be specific, we design
a novel power allocation (PA) strategy for balancing the fundamental
performance tradeoff between the S$\&$C spatial DoFs. We assume that
there always exists a LoS link for target detection or tracking. Since
unknown reflection coefficients may reduce the received signal energy
to a level that does not allow reliable
detection~\cite{fishler2006spatial}, the ISAC-BS estimates the
reflection coefficient of each path first, and then judiciously share
its total power across different paths through tailormade transmit
beamforming (TBF). To gain deeper insights into the PA design, we
first consider the special case, where there is only a single NLoS
link in addition to its LoS counterpart, and derive the optimal PA
scheme in closed form. Then, we extend it to a multiple-NLoS-link
scenario and provide a sub-optimal solution based on the popular
successive convex approximation (SCA) algorithm. Finally, our
simulations characterize the performance tradeoff between S$\&$C,
which indicates that both the S$\&$C performance can be simultaneously
optimized by effectively exploiting all ISAC spatial DoFs.

\section{System Model}
\begin{figure}[h]
    \centering
    \epsfxsize=1\linewidth
    \includegraphics[width=8cm]{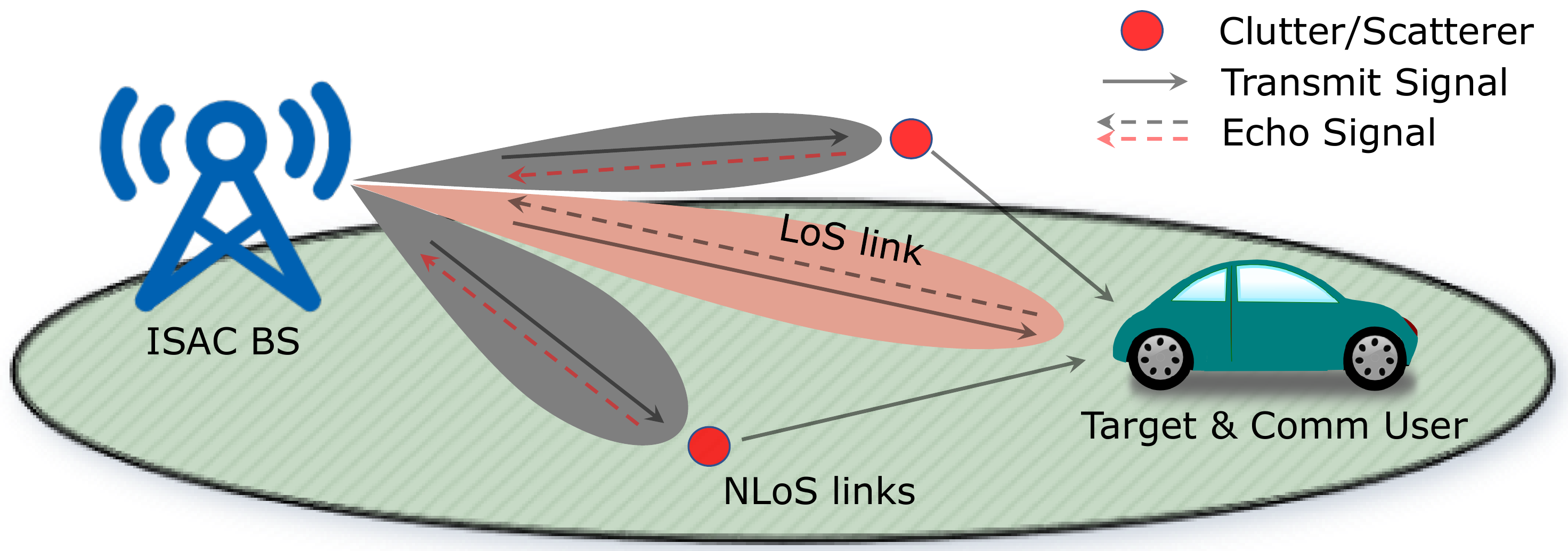}
    \caption{Illustration of the system considered.}
    \label{Fig1}
\end{figure}

We consider the downlink (DL) of a ISAC system, where the ISAC-BS is
equipped with $N_T$ transmit antennas (TAs) and $N_R$ receive antennas
(RAs) as shown in Fig. \ref{Fig1}. The ISAC-BS is serving a
single-antenna user (which is also treated as a point-like target) for
simultaneously supporting S$\&$C services. We assume that there are
$K$ channel impulse response (CIP) paths between the ISAC-BS and the
target, only one of which is the LoS link. Let $\mathcal{K}\triangleq
\{0,1,\cdots,K-1\}$ denote the index set of the $K$ propagation
paths. For notational convenience, we do not distinguish between
scatterers within each propagation path and clutter sources. As a
consequence, the ISAC-BS receives signal reflections from multiple
paths, where only the LoS path contains the target information of
interest, and the echoes from the NLoS paths are treated as
clutter. In what follows, we elaborate on the S$\&$C signal models.
\vspace{-0.4cm}  
\vspace{-0.5cm}  
\subsection{Sensing Signal Model}
Let $\mathbf{s}=\left[s_{0}, s_{1}, \ldots, s_{K-1}\right]^{T} \in \mathbb{C}^{K \times 1}$ denote the ISAC transmit signal vector. Thus the echo signal arriving at the ISAC-BS receiver can be expressed as 
\begin{align}
    \label{sensing signal}
\mathbf{y}_{r}=&\underbrace{{\beta}_{0} \sqrt{p_{0}} \mathbf{b}\left(\theta_{0}\right) \mathbf{a}^{H}\left(\theta_{0}\right) \mathbf{F} \mathbf{s}}_{\text {Target/LoS link }}+\underbrace{\sum_{k=1}^{K-1} {\beta}_{k} \sqrt{p_{k}}  \mathbf{b}\left(\theta_{k}\right) \mathbf{a}^{H}\left(\theta_{k}\right) \mathbf{F} \mathbf{s}}_{\text {Path-dependent Clutter/NLoS links}}\nonumber\\
&+\mathbf{z}_{r},
\end{align}
where $\beta_0$ and $\{\beta_k\}^{K-1}_{k=1}$ are independent and identically distributed (iid) reflection coefficients of the target and the $k$-th clutter/scatter source, while $p_0$ and $p_k$ represent the transmit power to the LoS and NLoS links, $\theta_0$ and $\theta_k$ are the angles of the target and the $k$-th path-dependent clutter source, $\mathbf{a}(\theta) =\frac{1}{\sqrt{N_{T}}}\left[1, e^{-j \pi \sin \theta}, \cdots, e^{-j \pi\left(N_{T}-1\right) \sin \theta}\right]^{T}$ and $\mathbf{b}(\theta) = \frac{1}{\sqrt{N_{R}}}\left[1, e^{-j \pi \sin \theta}, \cdots, e^{-j \pi\left(N_{R}-1\right) \sin \theta}\right]^{T}$ are the transmit and receive steering vectors, $ \mathbf{F}=\left[\mathbf{f}_{0}, \mathbf{f}_{1}, \ldots, \mathbf{f}_{K-1}\right] \in \mathbb{C}^{N_{T} \times K}$ denotes the TBF matrix, and $\mathbf{z}_{\mathbf{r}} \in \mathbb{C}^{N_{R} \times 1}$ represents the additive white Gaussian noise (AWGN) with variance of $\sigma^{2}_{R}$, respectively. 

Following the standard assumption of the radar
literature~\cite{fishler2006spatial,xu2008target,cui2013mimo}, and
given the fact that we focus our attention on power sharing among
multiple paths, we assume that the angle for each path is perfectly
predicted/tracked, thereby the beamforming gain
$\mathbf{a}^{H}(\theta_{k})\mathbf{f}_{k}$ equals to 1 and
$\mathbf{a}^{H}(\theta_{k})\mathbf{F}\mathbf{s}=s_{k}$. After receive
beamforming (RBF) at the ISAC-BS receiver, the sensing signal output
of the receiving filter is given by
\begin{align}
    \label{filter signal}
    y_{s} &=\beta_{0} \sqrt{p_{0}} \mathbf{w}^{H} \mathbf{b}\left(\theta_{0}\right) s_0 +\sum_{k=1}^{K-1} \beta_{k} \sqrt{p_{k}} \mathbf{w}^{H} \mathbf{b}\left(\theta_{k}\right) s_k + \mathbf{w}^{H}{\mathbf{z}_{r}},
\end{align}
where $\mathbf{w}$ is the RBF vector designed for maximizing the signal-to-clutter-plus-noise ratio (SCNR). As a consequence, the SCNR can be written as
\vspace{-0.5em}
\begin{align}
    \label{SCNR}
    \operatorname{SCNR}
    &=\frac{\left|s_{0} \beta_{0} \sqrt{p_{0}} \mathbf{w}^{H} \mathbf{b}\left(\theta_{0}\right)\right|^{2}}{\sum_{k=1}^{K-1}\left|s_{k} \beta_{k} \sqrt{p_{k}} \mathbf{w}^{H} \mathbf{b}\left(\theta_{k}\right)\right|^{2}+\mathbf{w}^{H} \mathbf{w} \sigma_{R}^{2}}\nonumber\\
    &=\frac{\frac{p_{0}\left|s_{0} \beta_{0}\right|^{2}}{\sigma_{R}^{2}}\left|\mathbf{w}^{H} \mathbf{b}\left(\theta_{0}\right)\right|^{2}}{\mathbf{w}^{H}\left(\boldsymbol{\Sigma} +\mathbf{I}_{N_{R}}\right) \mathbf{w}},
\end{align}
where $\boldsymbol{\Sigma}=\sum_{k=1}^{K-1} \frac{p_{k}\left|s_{k} \beta_{k}\right|^{2}}{\sigma_{R}^{2}} \mathbf{b}\left(\theta_{k}\right) \mathbf{b}^{H}\left(\theta_{k}\right)$ and $\mathbf{I}_{N_{R}}$ is the $N_R$-dimensional identity matrix. We note that the NLoS/clutter components are regarded as interference and thus they are present in the denominator.We also assume that each transmit symbol $s_k$ has unit power. The above SCNR maximization problem with respect to $\mathbf{w}$ is known as the minimum variance distortionless response (MVDR) beamforming problem~\cite{cui2013mimo}, which admits the closed-form solution of
\begin{align}
    \label{optimal w}
    \mathbf{w}^{\star}= \frac{\left[\boldsymbol{\Sigma}+\mathbf{I}_{N_{R}}\right]^{-1} \mathbf{b}\left(\theta_{0}\right)}{ \mathbf{b}^{H}\left(\theta_{0}\right)\left[\boldsymbol{\Sigma}+\mathbf{I}_{N_{R}}\right]^{-1} \mathbf{b}\left(\theta_{0}\right)}.
\end{align}

By substituting $\mathbf{w}^{\star}$ of (\ref{optimal w}) into (\ref{SCNR}), the maximum achievable SCNR can be expressed in the form of
\begin{align}
    \label{optimal SCNR}
    \operatorname{SCNR}=\frac{p_{0}\left|\beta_{0}\right|^{2}}{\sigma_{R}^{2}} \mathbf{b}^{H}\left(\theta_{0}\right)\left[\boldsymbol{\Sigma}+\mathbf{I}_{N_{R}}\right]^{-1} \mathbf{b}\left(\theta_{0}\right).
\end{align}

\vspace{-0.8em}
\subsection{Communication Signal Model}
We assume that the CIR is perfectly
known~\cite{liu2020Sensing-assistedComms,su2022secure}. The received
communication signal composed with LoS and NLoS components at the
target (which is also a communication user) is expressed as
\vspace{-0.5em}
\begin{align}
    \label{communication signal}
    y_{c} = \mathbf{h}^{H}\mathbf{Fs}+z_{c},   
\end{align}
where $\mathbf{h} \in \mathbb{C}^{N_T \times 1}$ represents the multiple-input-single-output (MISO) channel vector and $z_{c}$ is the AWGN at the target with a variance of $\sigma^2_{C}$. Based on the Rician fading model, we have 
\begin{align}
    \label{channel vector h }
    \mathbf{h} =\sqrt{\frac{\varrho }{1+\varrho }} \mathbf{h}_{\mathrm{LoS}} + \sqrt{\frac{1 }{1+\varrho }} \mathbf{h}_{\mathrm{NLoS}}, 
\end{align}
where $\varrho $ is the Rician factor and $\mathbf{h}_{\mathrm{LoS}} = \sqrt{p_{0}}\sqrt{N_{T}}\mathbf{a}\left(\theta_{0}\right)$ denotes the LoS component. The $K-1$ NLoS scattered components with may be expressed as $\mathbf{h}_{\mathrm{NLoS}} =  \sum_{k=1}^{K-1} \sqrt{p_{k}} \sqrt\frac{N_{T}} {K-1}\alpha_{k} \mathbf{a}\left(\theta_{k}\right)$, where $\alpha_{k} \thicksim \mathcal{CN}(0,1)$ is the complex path gain. 
Based on (\ref{communication signal}), the communication signal-to-noise ratio (SNR) is written as 
\vspace{-0.6em}
\begin{align}
    \label{SNR}
    \operatorname{SNR}=\frac{\left|\mathbf{h}^{H}\mathbf{Fs}\right|^2}{\sigma^2_{c}}=\frac{\left| \sum_{k = 0}^{K-1}\sqrt{p_k}\tilde{x}_k\right|^{2}}{\sigma_{c}^{2}},
\end{align}
where $\tilde{x}_0 = \sqrt{ \frac{N_T\varrho}{1+\varrho}}s_0$ and $\{\tilde{x}_k\}^{K-1}_{k=1} =  \sqrt{ \frac{N_T}{(K-1)(1+\varrho)}}\alpha_k s_k$ .

\subsection{Parameter Estimation}
Since all $\{\beta_k\}^{K-1}_{k=0}$ coefficients are unknown but iid,
the ISAC-BS has to estimate $\{\beta_k\}^{K-1}_{k=0}$ from different
paths and then judiciously allocate the transmit power based on all
estimated parameters in the first epoch, while the ISAC-BS performs
target detection in the second epoch. Again, we assume that
$\{\beta_k\}^{K-1}_{k=0}$ are iid; and subject to
$\mathcal{CN}(0,\sigma^2)$~\cite{fishler2006spatial}. Then, based on
\eqref{sensing signal}, the corresponding Linear Bayesian Estimation
model at sample $n$ is given by
\begin{align}
    \label{Bayesian Estimation Model}
\mathbf{y}_{r}[n]= \mathbf{H}[n]\boldsymbol{\beta}+\mathbf{z}_r[n], n = 1\ldots N,
\end{align}
where the $k$-th column of $\mathbf{H}[n]$ is
$\sqrt{p}s_k[n]\mathbf{b}\left(\theta_{k}\right)$, while $p = P_T / K$
denotes the initial transmit power used for estimation, and
$[\boldsymbol{\beta}]_k = \beta_k$ represents the parameters to be
estimated. Finally, $N$ is the number of sampling snapshots,
respectively. By stacking the signal snapshots observed into a single
vector, the estimation model becomes
\begin{align}
    \label{Completed Bayesian Estimation Model}
\mathbf{y} = \mathbf{H}\boldsymbol{\beta}+\mathbf{z}, 
\end{align}
\begin{align}
    \mathrm{where}~\mathbf{y} = \left[\begin{array}{c}
    \mathbf{y}_{r}\left[1\right]\\
    \mathbf{y}_{r}\left[2\right]\\
    \vdots  \\
    \mathbf{y}_{r}\left[{\tiny N}\right]\end{array}\right],
\mathbf{H} = \left[\begin{array}{c}
        \mathbf{H}\left[1\right]\\
        \mathbf{H}\left[2\right]\\
        \vdots  \\
        \mathbf{H}\left[{\tiny N}\right]\end{array}\right],
\mathbf{z} = \left[\begin{array}{c}
            \mathbf{z}_{r}\left[1\right]\\
            \mathbf{z}_{r}\left[2\right]\\
            \vdots  \\
            \mathbf{z}_{r}\left[{\tiny N}\right]\end{array}\right].\nonumber
\end{align} 

Note that $\boldsymbol{\beta}\sim
\mathcal{CN}(\mathbf{0},\sigma^2\mathbf{I}_{K})$ and $\mathbf{z}\sim
\mathcal{CN}(\mathbf{0},\sigma^{2}_{R}\mathbf{I}_{NN_R})$. Therefore,
the minimum mean square error (MMSE) estimator of $\boldsymbol{\beta}$
can be readily constructed as~\cite{kay1993fundamentals}
\begin{align}
    \label{Bayesian Estimator}
    \hat{\boldsymbol{\beta}} = \left(\frac{\sigma_R^{2}}{\sigma^{2}}\mathbf{I}_{K} + \mathbf{H}^H \mathbf{H} \right)^{-1}\mathbf{H}^H \mathbf{y}.
\end{align}
With the estimate \eqref{Bayesian Estimator} at hand, we approximate the SCNR as
\begin{align}
    \label{Estimater SCNR}
    \operatorname{SCNR}_{\mathrm{est}}=p_{0}\gamma_0\mathbf{b}^{H}\left(\theta_{0}\right)\left[\boldsymbol{\Sigma}_{\mathrm{est}}+\mathbf{I}_{N_{R}}\right]^{-1} \mathbf{b}\left(\theta_{0}\right),
\end{align}
where $\boldsymbol{\Sigma}_{\mathrm{est}}= \sum_{k=1}^{K-1}
p_{k}\gamma_k\mathbf{b}\left(\theta_{k}\right)
\mathbf{b}^{H}\left(\theta_{k}\right)$ and $\gamma_k =
\frac{\left|\hat{\beta}_{k}\right|^{2}}{\sigma_{R}^{2}}, k \in
\mathcal{K}$.  Since random fading may reduce the received signal
energy~\cite{fishler2006spatial}, We first have to identify whether
the target is present or absent, as detailed in the following.
\vspace{-0.5cm}  
\subsection{Target Detection}
We now proceed by constructing a hypothesis test, where we seek to choose between two hypotheses, i.e. $\mathcal{H} _1$, target present, or $\mathcal{H} _0$, target absent. This can be expressed as 
\begin{align}
    {y} =\left\{\begin{array}{l}
        \mathcal{H}_{0}: \sum_{k = 1}^{K-1}y_{k} + z, \\
        \mathcal{H}_{1}: y_0 + \sum_{k = 1}^{K-1}y_{k}  + z,
        \end{array}\right.
\end{align}
where $y_k =
\frac{\sqrt{p_k}{\beta}_k\mathbf{w}^{H}\mathbf{b}\left(\theta_{k}\right)}{\sqrt{N}}\sum_{n
  = 1}^{N}s_k[n]s^{\ast }_0[n], k \in \mathcal{K}$,
$z=\frac{1}{\sqrt{N}}\sum_{n =
  1}^{N}\mathbf{w}^{H}\mathbf{z}_r[n]s^{\ast }_0[n]$ and
$\frac{1}{\sqrt{N}}s^{\ast }_0[n]$ is the matching signal. Here
$1/\sqrt{N}$ is set for ensuring that the received signal energy
remains constant after matched filtering. By noting that
$\beta_k\thicksim \mathcal{CN}(0,\sigma^2)$ and $z\thicksim
\mathcal{CN}(0, \|\mathbf{w}\Vert ^2\sigma^{2}_{R})$, we have ${y}
|\mathcal{H}_{0} \thicksim \mathcal{CN}(0,\eta_0)$ and ${y}
|\mathcal{H}_{1} \thicksim \mathcal{CN}(0,\eta_1)$, where $\eta_0=
\sum_{k =
  1}^{K-1}{p_k}|\mathbf{w}^H\mathbf{b}\left(\theta_{k}\right)|^2\sigma^2
+ \|\mathbf{w}\Vert ^2\sigma^{2}_{R}$ and $\eta_1= \sum_{k =
  0}^{K-1}{p_k}|\mathbf{w}^H\mathbf{b}\left(\theta_{k}\right)|^2\sigma^2
+ \|\mathbf{w}\Vert ^2\sigma^{2}_{R}$. Based on the above, we can now
formulate the Neyman-Pearson detector~\cite{kay1993fundamentals}
\begin{align}
    T = |y| ^2 \underset{\mathcal{H}_{0}}{\overset{\mathcal{H}_{1}}\gtrless} \delta,
\end{align}
in which the threshold $\delta$ is set to satisfy the maximum tolerant
probability of false alarm $P_{\mathrm{FA}}$. Thus the value of $T$ is
distributed as $ T|\mathcal{H}_{0} \thicksim \frac{\eta_0}{2}
\chi_{2}^{2} $ and $ T|\mathcal{H}_{1} \thicksim \frac{\eta_1}{2}
\chi_{2}^{2}$, respectively, where $\chi_{2}^{2}$ denotes the central
chi-squared distribution having a DoF $=2$.

Given a constant $P_{\mathrm{FA}}$, it follows that $\delta$ can be
set to~\cite{fishler2006spatial} $\delta = \frac{\eta_0}{2}
F_{\chi_{2}^{2}}^{-1}\left(1-P_{\mathrm{FA}}\right)$, where
$F_{\chi_{2}^{2}}^{-1}$ denotes the inverse cumulative distribution
function of the chi-square distribution. Accordingly, the probability
of correct detection $P_D$ is~\cite{fishler2006spatial}
\begin{align}
    P_{D}=\mathbb{P}(T>\delta|\mathcal{H}_{1})
    =1 - F_{\chi_{2}^{2}}\left( \frac{\eta_0}{\eta_1}F_{\chi_{2}^{2}}^{-1}\left(1-P_{\mathrm{FA}}\right)\right).
\end{align}

\section{Problem Formulation and solution}
In this section, we discuss the PA problem across multiple propagation
paths. Our aim is to exploit ISAC channel as DoFs tradeoff for
maximizing the communication performance, while still satisfying the
sensing performance.
\subsection{Optimization Problem Formulation}
It is provable that $P_D$ is monotonically decreasing with $\eta_0 /
\eta_1$, i.e., $P_D \varpropto\mathrm{SCNR}$, both of which are
determined by the specific PA. Therefore, to constrain $P_D$ is
equivalent to constraining the SCNR. In what follows, we choose SCNR
as our sensing performance metric. By employing the achievable
communication rate as our objective function, the PA problem can be
formulated as
\begin{subequations}\label{P1}
\begin{align}
    \max_{ \mathbf{p} } &~~ \log_2(1+\mathrm{SNR})\\
    ~~~\mathrm{s.t.} &~~ \operatorname{SCNR}_{\mathrm{est}} \ge \Gamma , \\
    \label{power >0}
    &~~\mathbf{1}^{T}\mathbf{p} \leq  P_{T}, p_{k} \ge 0,  \forall k \in \mathcal{K},
\end{align}
\end{subequations}
where $\mathbf{p} = [p_0, p_1,\ldots, p_{K-1}]^{T}$ is the PA vector
to be optimized, $\Gamma > 0$ is the minimum required SCNR, and $P_T$
is the total transmit power budget. It can be readily seen that
equality holds for $\mathbf{1}^{T}\mathbf{p} \leq P_{T}$, when the
optimum is reached. By noting \eqref{SNR}, problem \eqref{P1} can be
equivalently recast as
\begin{subequations}\label{P2}
\begin{align}
    \max_{ \mathbf{p}} &~~ {\left| \sum_{k = 0}^{K-1}\sqrt{p_k}\tilde{x}_k\right|^{2}} \\
    \label{SCNR constraint}
    ~~~\mathrm{s.t.} &~~ \operatorname{SCNR}_{\mathrm{est}} \ge \Gamma , \\
    \label{power constraint}
    &~~\mathbf{1}^{T}\mathbf{p} =  P_{T}, p_{k} \ge 0,  \forall k \in \mathcal{K}.
\end{align}
\end{subequations}

Since we assume that the communication CIR is perfectly known at the ISAC-BS~\cite{liu2020Sensing-assistedComms,su2022secure}, one can simply compensate for both the phases of the complex path gain factor $\alpha_k$ and the complex transmit signals $s_k$ via the following modification of the TBF matrix
\begin{align}
    \label{precoding matrix}
    \mathbf{F}=\left[\mathbf{f}_{0}e^{-jg_0}, \mathbf{f}_{1}e^{-jg_1}, \ldots, \mathbf{f}_{K-1}e^{-jg_{K-1}}\right],
\end{align}
where $g_0 = \angle s_0$ represents the phase of the complex transmit
symbol in the LoS link, and $g_k = \angle \alpha_k s_k, k = 1, \ldots,
K-1,$ denotes the phase of the complex path gain factor and the
complex transmit symbol in the $k$-th NLoS link, respectively. Based
on \eqref{precoding matrix}, we have
$\mathbf{a}^{H}\left(\theta_{0}\right) \mathbf{F} \mathbf{s} = |s_0|$
and $ \alpha_k\mathbf{a}^{H}\left(\theta_{0}\right) \mathbf{F}
\mathbf{s} = |\alpha_k s_k|, k = 1, \ldots, K-1$. We remark that the
modified TBF matrix $\mathbf{F}$ does not affect the radar
SCNR. Recalling the fact that each symbol has unit power, we can
reformulate problem \eqref{P2} as
\begin{align}\label{P3}
    \max_{ \mathbf{p}} ~~ { \sum_{k = 0}^{K-1}\sqrt{p_k}x_k} 
    ~~~\mathrm{s.t.} ~~  \eqref{SCNR constraint}~\mathrm{and}~\eqref{power constraint},
\end{align}
where $x_0 =\sqrt{ \frac{N_T\varrho}{1+\varrho}}$ and $\{x_k\}^{K-1}_{k=1} = \sqrt{ \frac{N_T}{(K-1)(1+\varrho)}}|{\alpha_k}|$.

However, problem~\eqref{P3} is still non-convex, since \eqref{SCNR
  constraint}, which is expressed as a convex function greater than a
given threshold, is non-convex. This makes it challenging for us to
design an efficient solver for \eqref{P3}. More particularly, dealing
with the non-convex constraint \eqref{SCNR constraint} is quite
challenging. To gain deeper insights here, we consider a pair of cases
having different numbers of NLoS propagation paths, a.k.a., clutter
sources. Specifically, a single NLoS link ($K=2$) and multiple NLoS
links ($K>2$) are considered.
\vspace{-0.5em}
\subsection{Single NLoS Link Case}
Firstly, we assume that there are only two channel links, i.e. CIR
taps between the ISAC-BS and the target. This special case corresponds
to the V2I scenario, where the signals received at the vehicle come
from both the direct LoS link and a ground-reflected NLoS link. In
this case, problem \eqref{P3} can be simplified to
\begin{subequations}\label{One_NLoS_Model}
\begin{align}
    \max_{ \{p_0,p_1\} } &~~ \sqrt{p_{0}}x_{0}+\sqrt{p_{1}}x_{1} \\
    \label{SCNR cons case1}
    ~~~\mathrm{s.t.} &~~ 
    p_{0}\gamma_0 \mathbf{b}^{H}\left(\theta_{0}\right)\left[\boldsymbol{\Sigma}\left(p_{1}\right)+\mathbf{I}_{N_{R}}\right]^{-1} \mathbf{b}\left(\theta_{0}\right) \ge \Gamma, \\
    \label{power constraints}
    &~~ p_{0}+p_{1}=P_{T},p_{0} \ge 0, p_{1} \ge 0,
\end{align}
\end{subequations}
where
$\boldsymbol{\Sigma}(p_{1})=p_{1}\gamma_1\mathbf{b}\left(\theta_{1}\right)
\mathbf{b}^{H}\left(\theta_{1}\right)$. Note that
\eqref{One_NLoS_Model} is a non-convex optimization problem which is
difficult to solve directly. Thanks to Woodbury's matrix
identity~\cite{kay1993fundamentals}, we have
\begin{align}\label{Inverse_Formula}
    \left[\boldsymbol{\Sigma}\left(p_{1}\right)+\mathbf{I}_{N_{R}}\right]^{-1} = \mathbf{I}_{N_{R}} - \frac{p_1 \gamma_1}{p_1 \gamma_1 + 1}\mathbf{b}\left(\theta_{1}\right) \mathbf{b}^{H}\left(\theta_{1}\right).
\end{align}

Then, by substituting \eqref{power constraints} and \eqref{Inverse_Formula} into \eqref{SCNR cons case1}, we have
\begin{align}
    \label{h(p_1)}
    h(p_1) \triangleq  Ap_1^2 + Bp_1 + C \le  0,
\end{align}
where $A = (1-b)\gamma_0 \gamma_1$, $B = \Gamma \gamma_1 + \gamma_0 +
P_T(b-1)\gamma_0 \gamma_1$, $C = \Gamma - P_T \gamma_0$, $b =
\mathbf{b}^{H}\left(\theta_{0}\right)\mathbf{b}\left(\theta_{1}\right)\mathbf{b}^{H}\left(\theta_{1}\right)\mathbf{b}\left(\theta_{0}\right)$. After
the above derivation, we have the following equivalent constraint as
for \eqref{SCNR cons case1} and \eqref{power constraints}
\begin{align}
    \label{h(p_1) equivalent constraint}
    0 \leq p_1 \leq \min \left\{P_A, P_T\right\}, 
\end{align}
where $P_A$ is the positive root of $h(p_1)=0$, whose expression is omitted here. Thus, problem \eqref{One_NLoS_Model} can be reformulated as
\begin{align}\label{Function Programming}
    \max_{ p_1 } ~~ f(p_1) ~~\mathrm{s.t.} ~~ 0 \leq p_1 \leq \min \left\{P_A, P_T\right\}, 
\end{align}
where we have $f(p_1) = \sqrt{P_T-p_{1}}x_0+\sqrt{p_{1}}x_1$. It is
not difficult to show that (P\ref{Function Programming}) boils down to
a one-dimensional optimization problem associated with a given
feasible interval. The optimal solution $p^{\star}_1$ can be readily
expressed as:
\begin{align}
    p^{\star}_1 = \argmax_{\mathcal{D}}~f(p_1),
\end{align}
where $\mathcal{D} = \left\{ \min \left\{{ P_A, {P_B}}\right\},
0\right\}$ denotes the set of boundary points and extreme point of the
objective function. Here the extreme point $P_B = \frac{P_{T}
  {x}_{1}^{2}}{{x}_{0}^{2}+{x}_{1}^{2}}$ can be calculated by solving
$f^{\prime }(p_1)= \frac{-\frac{1}{2}
  {x}_{0}}{\sqrt{P_{T}-p_{1}}}+\frac{\frac{1}{2}
  {x}_{1}}{\sqrt{p_{1}}} = 0.$

\subsection{Multiple NLoS Links}
Now we investigate the multiple-NLoS-link scenario by tackling the
non-convex constraint \eqref{SCNR constraint} in \eqref{P3}. For
notational convenience, we let
$g(\mathbf{p}_c)$$=\mathbf{b}^{H}\left(\theta_{0}\right)\left[\boldsymbol{\Sigma}_{\mathrm{est}}+\mathbf{I}_{N_{R}}\right]^{-1}
\mathbf{b}\left(\theta_{0}\right)$, where $\mathbf{p}_c = [p_1,
  p_2,\ldots, p_{K-1}]^{T} \in \mathbb{R}^{K-1}$ represents the
transmit power vector with each element respectively the power
allocated to each NLoS link. It is plausible that $g(\mathbf{p}_c)$ is
convex with respect to $\mathbf{p}_c$ by its convex
Epigraph~\cite{boyd2004convex}.

Observe that any convex function is globally lower-bounded by its
first-order Taylor expansion at any point~\cite{boyd2004convex}, thus
the lower bound of $g(\mathbf{p}_c)$ is formulated as
\begin{align}\label{SCA Phi}
    g(\mathbf{p}_c) \ge  g_{lb}(\mathbf{p}_c)\triangleq g(\mathbf{p}^r_c) +(\nabla g(\mathbf{p}^r_c))^T(\mathbf{p}_c - \mathbf{p}^r_c),
\end{align}
where $\mathbf{p}^r_c$ is the point given at the $r$-th iteration and
$\nabla g(\mathbf{p}^r_c)$ is the gradient vector at
$\mathbf{p}^r_c$. The $k$-th element of $\nabla g(\mathbf{p}^r_c)$ is
$-\mathbf{b}^{H}\left(\theta_{0}\right)\left[\boldsymbol{\Sigma}_{\mathrm{est}}^r+\mathbf{I}_{N_{R}}\right]^{-1}
\mathbf{B}_k
\left[\boldsymbol{\Sigma}_{\mathrm{est}}^r+\mathbf{I}_{N_{R}}\right]^{-1}
\mathbf{b}\left(\theta_{0}\right)$, where
$\boldsymbol{\Sigma}_{\mathrm{est}}^r = \sum_{k=1}^{K-1}
p_{k}^r\gamma_k\mathbf{b}\left(\theta_{k}\right)
\mathbf{b}^{H}\left(\theta_{k}\right)$ is a constant matrix at the
$r$-th iteration and
$\mathbf{B}_k=\mathbf{b}\left(\theta_{k}\right)\mathbf{b}^{H}\left(\theta_{k}\right)$. By
introducing the SCA technique, we can thus formulate a sub-problem in
each iteration as
\vspace{-0.8em}
\begin{align}\label{P4}
    \max_{ p_0, \mathbf{p}_c} &~~ { \sum_{k = 0}^{K-1}\sqrt{p_k}x_k}\\ 
    ~~~\mathrm{s.t.} &~~  g_{lb}(\mathbf{p}_c) - \frac{\Gamma}{\gamma_0{p}_0 }\ge 0,~\mathrm{and}~\eqref{power constraint}.\nonumber
\end{align}
Observe that problem \eqref{P4} is convex and thus can be solved by CVX directly\cite{boyd2004convex}. Therefore, the original optimization problem \eqref{P3} can be solved in an iterative manner, which is characterized in \textbf{Algorithm} \ref{alg1}.
\begin{algorithm}[h]
    \caption{SCA Based Power Allocation.}
    \label{alg1}
    \begin{algorithmic}[1]
    \Require
    $P_T, \alpha_k, \sigma^{2}_{R} , \sigma^{2}_{C}, \theta_0, \theta_k, N_T,N_R, K,r_{\mathrm{max}},\epsilon$.
    \Ensure
    $\mathbf{p}$.
    \State Initialize the transmit power $\mathbf{p}^r$, r = 1.
    \State Initialize the estimated reflecting coefficients by \eqref{Bayesian Estimator}.
    \Repeat
    \State Calculate $\nabla g(\mathbf{p}^r_c)$.
    \State Solve problem \eqref{P4} for given $\mathbf{p}^r$. Denote the optimal solution as $\mathbf{p}^{r+1}$.
    \State Update $r=r+1$.
    \Until The increase of the objective value is below $\epsilon=10^{-5}$ or $r=r_{\mathrm{max}}$.
    \end{algorithmic}
\end{algorithm}
\vspace{-1em}

\vspace{-0.5em}
\begin{figure*}[t]
		\begin{minipage}[t]{0.3\linewidth}
			\centering
			\includegraphics[width=2.4in]{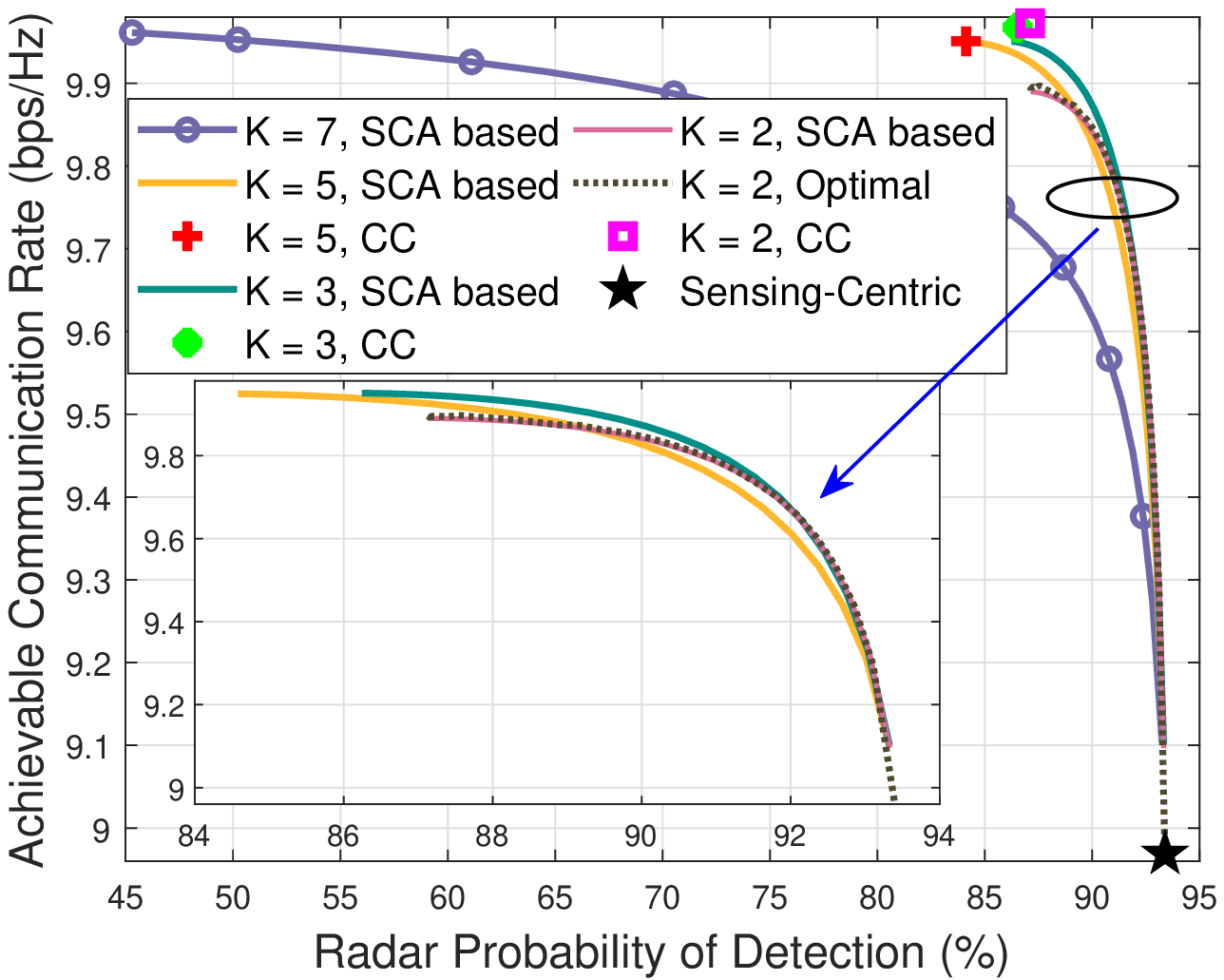}
			\caption{S$\&$C performance tradeoff.}
			\label{Performance}
	\end{minipage}
	\hspace{0.01\linewidth}	
		\begin{minipage}[t]{0.3\linewidth}
			\centering
			\includegraphics[width=2.4in]{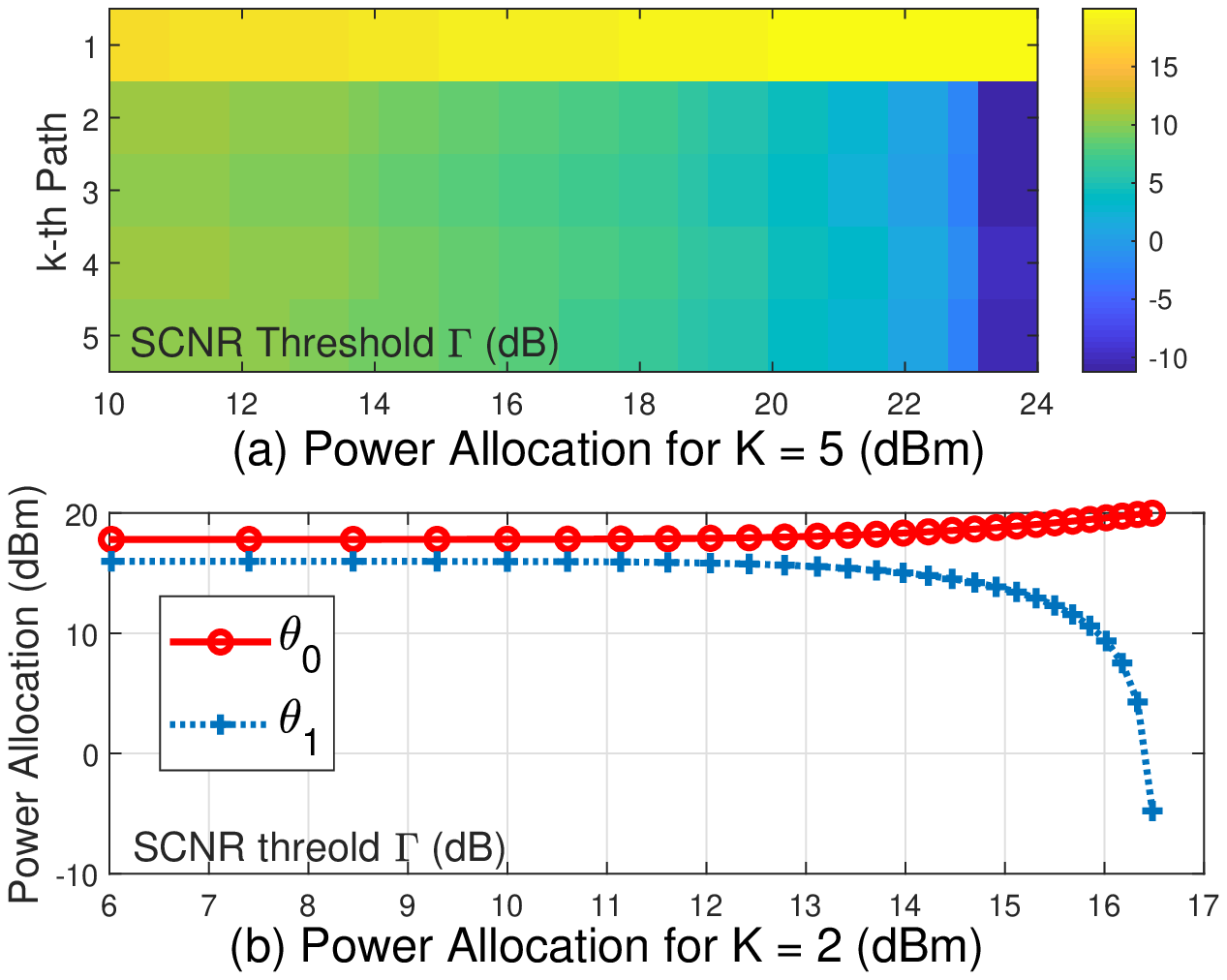}
			\caption{PA with $K = 5$ and $K = 2$.}
			\label{Power Allocation over K}
	\end{minipage}
	\hspace{0.01\linewidth} 
		\begin{minipage}[t]{0.3\linewidth}
			\centering
			\includegraphics[width=2.4in]{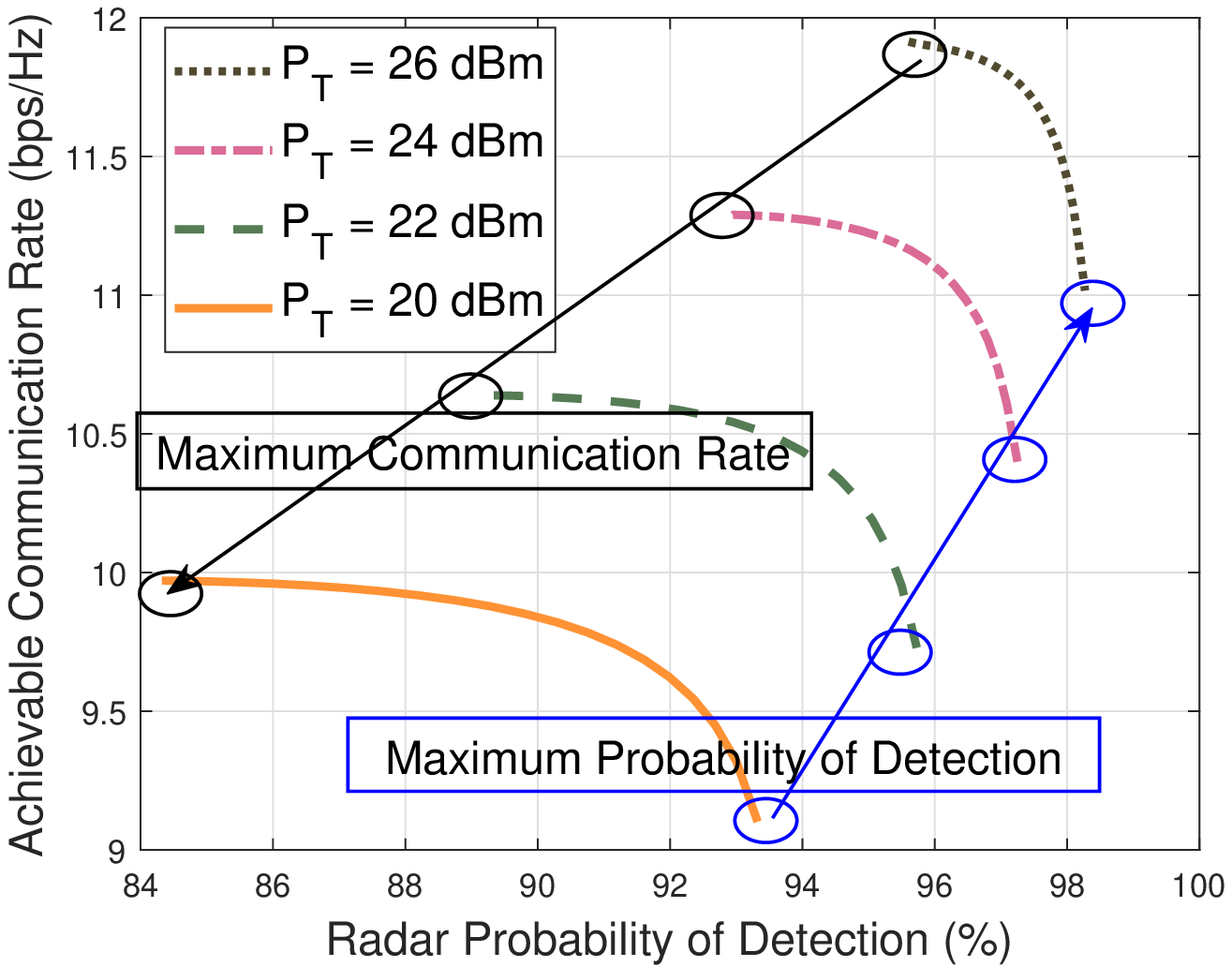}
			\caption{S$\&$C performance versus $P_T$.}
			\label{Different P_T}
	\end{minipage}
\end{figure*}
\vspace{-0.5em}
\section{Simulation Results}
In this section, we evaluate the proposed algorithm by MonteCarlo
based simulation results. Without loss of generality, the reflecting
coefficients $\beta_k$, and the complex channel gain $\alpha_k$ are
assumed to obey the standard Complex Gaussian distribution
$\mathcal{CN}(0,1)$ and the Rician factor is set as $\varrho =
1$. Unless otherwise specified, the maximum transmit power is given as
$P_T = 20$ dBm and the noise power is set as $\sigma_c^2 = \sigma_r^2
= 0$ dBm. The target is located at $\theta_0 = 0^{\circ}$ (LoS link)
and the clutter sources (NLoS links) are located at $\theta_1 =
-20^{\circ}, \theta_2 = -10^{\circ}, \theta_3 = 10^{\circ}, \theta_4 =
20^{\circ}$, respectively. The SCNR threshold $\Gamma$ spans from $0 $
dB to $\Gamma_T$ dB, where $\Gamma_T = P_T\gamma_0$ is the maximum
SCNR threshold amputated for avoiding an infeasible case.

We commence by evaluating the S$\&$C performance under different DoFs in Fig. \ref{Performance} through the proposed approaches, using one pair of benchmarks, namely,
\begin{itemize}
    \item[$\bullet$] Sensing-centric (SC) design, which reaches the best sensing performance by setting $p_0 = P_T$, i.e., the ISAC-BS allocates all the power to the LoS link;
    \item[$\bullet$] Communication-centric (CC) design, which reaches the maximum achievable communication rate by applying the Cauchy-Schwarz inequality to the objective function of \eqref{P4}, i.e., $(\sum_{k = 0}^{K-1}\sqrt{p_k}x_k)^2 \leq P_T(\sum_{k = 0}^{K-1} x_k^2)$.
\end{itemize}

First, we observe that the SCA based method approaches the optimal
solution for $K = 2$, because the SCA technique reduces the feasible
region. Then, we also see that regardless of the spatial DoFs, the
achievable maximum radar probability of detection is identical for all
cases since the ISAC-BS assigns its total power to the LoS link. By
contrast, when the spatial DoFs are higher, our approach usually
attains a higher communication rate (see $K = 2$, $K = 3$, $K = 5$ and
$K = 7$ in Fig. \ref{Performance}). But this is not always true, when
we also take the sensing performance into consideration, especially
when the radar probability of detection has to be above $90\%$. This
is because having higher DoFs for communications imposes more clutter
sources, which are harmful to radar sensing. Futhermore, all proposed
schemes are capable of attaining a flexible CC or SC trade-off by
adjusting the PA among different paths. Again, CC design exploits all
available spatial DoFs, while SC design only relies on LoS
transmission. We can conclude through Fig. \ref{Performance} that
utilizing flexible spatial DoFs tradeoff is important for achieving
scalable S$\&$C performance.

In Fig. \ref{Power Allocation over K}, we characterize our PA scheme
for $K = 5$ and $K = 2$ respectively based \textbf{Algorithm}
\ref{alg1}. Observe that upon increasing the SCNR threshold, the
transmit power allocated to the NLoS links is reduced. When the SCNR
approaches $\Gamma_T$, i.e., the maximum feasibility threshold, the
transmit power allocated to the LoS link tends to approach $P_T$ for
satisfying more strict SCNR threshold requirements. Futhermore, since
we assume that both the reflecting coefficients and the complex
channel gain are iid, the transmit power of different NLoS links is
commensurate (see $K = 5$ in Fig. \ref{Power Allocation over K}(a)).

Finally, in Fig. \ref{Different P_T}, we show the S$\&$C performance
tradeoff vs. $P_T$, for $K = 5$. The maximum achievable communication
rate and the probability of successful detection increase with the
maximum transmit power budget. Moreover, the S$\&$C performance region
is also extended upon increasing the power budget.
\section{Conclusions}
In this paper, we investigated the S$\&$C performance tradeoffs in
terms of the ISAC channel's DoFs, by proposing a novel transmit power
allocating accords the transmit signal propagation paths. We first
constructed the parameter estimator and target detector under the
signal model considered. Then, we formulated a PA problem for
maximizing the achievable communication rate subject to a minimal
required radar SCNR constraint and power budget. To gain deeper
insights into the problem, the closed-form solution was focused for
the case of a single NLoS link. Then, we extended it to the more
practical scenario of multiple NLoS links. To tackle the resultant
non-convex optimization problem, we harnessed the SCA algorithm, where
the transmit power vector is optimized in an iterative
manner. Finally, simulation results were provided for characterizing
the performance tradeoff between S$\&$C, which suggests that both the
S$\&$C performance can be simultaneously optimized by exploiting all
ISAC spatial DoFs.
\vspace{-0.5em}
\bibliographystyle{IEEEtran}
\bibliography{Ref_DoF}

\begin{thebibliography}{10}
\providecommand{\url}[1]{#1}
\csname url@samestyle\endcsname
\providecommand{\newblock}{\relax}
\providecommand{\bibinfo}[2]{#2}
\providecommand{\BIBentrySTDinterwordspacing}{\spaceskip=0pt\relax}
\providecommand{\BIBentryALTinterwordstretchfactor}{4}
\providecommand{\BIBentryALTinterwordspacing}{\spaceskip=\fontdimen2\font plus
\BIBentryALTinterwordstretchfactor\fontdimen3\font minus
  \fontdimen4\font\relax}
\providecommand{\BIBforeignlanguage}[2]{{%
\expandafter\ifx\csname l@#1\endcsname\relax
\typeout{** WARNING: IEEEtran.bst: No hyphenation pattern has been}%
\typeout{** loaded for the language `#1'. Using the pattern for}%
\typeout{** the default language instead.}%
\else
\language=\csname l@#1\endcsname
\fi
#2}}
\providecommand{\BIBdecl}{\relax}
\BIBdecl

\bibitem{cui2021integrating}
Y.~Cui, F.~Liu, X.~Jing, and J.~Mu, ``Integrating sensing and communications
  for ubiquitous {I}o{T}: Applications, trends, and challenges,'' \emph{IEEE
  Network}, vol.~35, no.~5, pp. 158--167, 2021.

\bibitem{liu2021integrated}
F.~Liu, Y.~Cui, C.~Masouros, J.~Xu, T.~X. Han, Y.~C. Eldar, and S.~Buzzi,
  ``Integrated sensing and communications: Towards dual-functional wireless
  networks for 6{G} and beyond,'' \emph{IEEE J. Sel. Areas Commun.}, pp. 1--1,
  early access, 2022, doi: {10.1109/JSAC.2022.3156632}.

\bibitem{liu2020Sensing-assistedComms}
F.~Liu, W.~Yuan, C.~Masouros, and J.~Yuan, ``Radar-assisted predictive
  beamforming for vehicular links: Communication served by sensing,''
  \emph{IEEE Trans. Wireless Commun.}, vol.~19, no.~11, pp. 7704--7719, 2020.

\bibitem{su2022secure}
N.~Su, F.~Liu, Z.~Wei, Y.-F. Liu, and C.~Masouros, ``Secure dual-functional
  radar-communication transmission: Exploiting interference for resilience
  against target eavesdropping,'' \emph{IEEE Trans. Wireless Commun.}, 2022.

\bibitem{li2021rethinking}
G.~Li, S.~Wang, J.~Li, R.~Wang, F.~Liu, M.~Zhang, X.~Peng, and T.~X. Han,
  ``Rethinking the tradeoff in integrated sensing and communication:
  Recognition accuracy versus communication rate,'' \emph{arXiv preprint
  arXiv:2107.09621}, 2021.

\bibitem{yang2020dual}
J.~Yang, G.~Cui, X.~Yu, and L.~Kong, ``Dual-use signal design for radar and
  communication via ambiguity function sidelobe control,'' \emph{IEEE Trans.
  Veh. Technol.}, vol.~69, no.~9, pp. 9781--9794, 2020.

\bibitem{zhang2022integrated}
R.~Zhang, B.~Shim, W.~Yuan, M.~Di~Renzo, X.~Dang, and W.~Wu, ``Integrated
  sensing and communication waveform design with sparse vector coding: Low
  sidelobes and ultra reliability,'' \emph{IEEE Trans. Veh. Technol.}, 2022.

\bibitem{wang2021joint}
X.~Wang, Z.~Fei, Z.~Zheng, and J.~Guo, ``Joint waveform design and passive
  beamforming for ris-assisted dual-functional radar-communication system,''
  \emph{IEEE Trans. Veh. Technol.}, vol.~70, no.~5, pp. 5131--5136, 2021.

\bibitem{fishler2006spatial}
E.~Fishler, A.~Haimovich, R.~S. Blum, L.~J. Cimini, D.~Chizhik, and R.~A.
  Valenzuela, ``Spatial diversity in radars---models and detection
  performance,'' \emph{IEEE Trans. Signal Process.}, vol.~54, no.~3, pp.
  823--838, 2006.

\bibitem{xu2008target}
L.~Xu, J.~Li, and P.~Stoica, ``Target detection and parameter estimation for
  {MIMO} radar systems,'' \emph{IEEE Trans. Aerosp. Electron. Syst.}, vol.~44,
  no.~3, pp. 927--939, 2008.

\bibitem{cui2013mimo}
G.~Cui, H.~Li, and M.~Rangaswamy, ``{MIMO} radar waveform design with constant
  modulus and similarity constraints,'' \emph{IEEE Trans. Signal Process.},
  vol.~62, no.~2, pp. 343--353, 2013.

\bibitem{liuxiang2020joint}
X.~Liu, T.~Huang, N.~Shlezinger, Y.~Liu, J.~Zhou, and Y.~C. Eldar, ``Joint
  transmit beamforming for multiuser {MIMO} communications and {MIMO} radar,''
  \emph{IEEE Trans. Signal Process.}, vol.~68, pp. 3929--3944, 2020.

\bibitem{kay1993fundamentals}
S.~M. Kay, \emph{Fundamentals of statistical signal processing: estimation
  theory}.\hskip 1em plus 0.5em minus 0.4em\relax Prentice-Hall, Inc., 1993.

\bibitem{boyd2004convex}
S.~Boyd, S.~P. Boyd, and L.~Vandenberghe, \emph{Convex optimization}.\hskip 1em
  plus 0.5em minus 0.4em\relax Cambridge university press, 2004.

\end{thebibliography}
\end{document}